\documentclass[preprint,review,11pt]{elsarticle}
\usepackage{url}
\usepackage{amsmath,amssymb} 
\usepackage{algorithm}
\usepackage{algpseudocode}
\usepackage{graphicx}
\usepackage{lineno}
\algnotext{EndFor}
\algnotext{EndIf}
\algnotext{EndFunction}
\newcommand{\vecr}{\mathbf{r}}
\newcommand{\vecx}{\mathbf{x}}
\biboptions{square,comma,numbers,sort&compress}
\journal{Computer Physics Communications}

\begin{document}
\begin{frontmatter}

\title{SAWdoubler: a program for counting self-avoiding walks}
\author[lei]{Raoul D. Schram}
\ead{Schram@lorentz.leidenuniv.nl}
\author[lei,uup]{Gerard T. Barkema}
\ead{g.t.barkema@uu.nl}
\author[uum]{Rob H. Bisseling\corref{cor1}}
\ead{r.h.bisseling@uu.nl}
\cortext[cor1]{Corresponding author}
\address[lei]{Instituut-Lorentz for Theoretical Physics, Leiden University,
P.O. Box 9504,\\ 2300 RA Leiden, The Netherlands}
\address[uup]{Institute for Theoretical Physics, Utrecht University,
P.O. Box 80195,\\ 3508 TD  Utrecht, The Netherlands}
\address[uum]{Mathematical Institute, Utrecht University,
P.O. Box 80010,\\ 3508 TA Utrecht, The Netherlands}

\begin{abstract}
This article presents SAWdoubler, a package for counting the total
number $Z_N$ of self-avoiding walks (SAWs) on a regular lattice by the
length-doubling method, of which the
basic concept has been published previously by us.
We discuss an algorithm for the creation of all SAWs of length $N$,
efficient storage of these SAWs in a tree data structure, and an algorithm
for the computation of correction terms to the count $Z_{2N}$ for
SAWs of double length, removing all combinations of two intersecting
single-length SAWs.

We present an efficient numbering of the lattice sites that enables
exploitation of symmetry and leads to a smaller tree data structure;
this numbering is by increasing Euclidean distance from the origin of
the lattice. Furthermore, we show how the computation can be parallelised
by distributing the iterations of the main loop of the algorithm
over the cores of a multicore architecture.
Experimental results on the 3D cubic lattice demonstrate that $Z_{28}$
can be computed on a dual-core PC in only 1 hour and 40 minutes,
with a speedup of 1.56 compared to the single-core computation and
with a gain by using symmetry of a factor of 26. We present results
for memory use and show how the computation is made to fit in 4
Gbyte RAM. It is easy to extend the SAWdoubler software to other
lattices; it is publicly available under the GNU LGPL license.
\end{abstract}


\end{frontmatter}

%\begin{linenumbers}

% Computer program descriptions should contain the following
% PROGRAM SUMMARY.

\textbf{Program Summary}

\begin{small}
\noindent
\emph{Manuscript title:} SAWdoubler: a program for counting self-avoiding walks\\
\emph{Authors:} Raoul D. Schram, Gerard T. Barkema, Rob H. Bisseling \\
\emph{Program title:} SAWdoubler \\
\emph{Journal reference:}     \\
  %Leave blank, supplied by Elsevier.
\emph{Catalogue identifier:}                                   \\
  %Leave blank, supplied by Elsevier.
\emph{ Program obtainable from:} CPC Program Library, Queen's University, Belfast,
N.Ireland; also from \url{http://www.staff.science.uu.nl/~bisse101/SAW/}\\
\emph{Number of lines of code of program:} 1152\\
\emph{Licensing provisions:} GNU LGPL  \\
\emph{Distribution format:} tar.gz\\
\emph{Programming language:} C  \\
\emph{Computer:} Any computer with a UNIX-like operating system and a 
C compiler. For large  problems, use is made of specific 128-bit integer
arithmetic provided by the gcc compiler. \\
\emph{Operating system:} Any UNIX-like system; developed under Linux and 
Mac OS 10\\
\emph{RAM:} Problem dependent (2 Gbyte for counting SAWs of length 28
on the 3D cubic lattice)                                          \\
\emph{Number of processors used:} 1. Parallel version available in directory Extras. 
\emph{Classification:} 7. Condensed matter and surface science \\
\emph{Nature of problem:}
Computing the number of self-avoiding walks of a given length on a given
lattice\\
\emph{Solution method:} Length doubling \\
\emph{Restrictions:} The length of the walk must be even.
Lattice is 3D simple cubic.  \\
\emph{Additional comments:} The lattice can be replaced by other lattices,
such as BCC, FCC, or a 2D square lattice\\
\emph{Running time:} Problem dependent (2.5 hours using one processor core
for length 28 on the 3D cubic lattice) \\
\end{small}

%maketitle
\section{Introduction}
Counting the number of self-avoiding walks on a regular lattice is 
a fundamental problem in combinatorics and statistical physics.
A \emph{self-avoiding walk} (SAW) is a path in a lattice where each step
goes from a lattice point to an adjacent point in the lattice,
and where a previously visited point cannot be visited again.
The SAW enumeration problem is of importance in physics because 
a SAW can be used to model the conformation of a polymer, where two monomers are forbidden
to occupy the same location (the excluded-volume principle).
Furthermore, this problem has been a challenge to mathematicians and physicists alike,
because counting the exact number of SAWs is difficult.
The number $Z_N$ of SAWs of length $N$ grows quickly with $N$, asymptotically as
\begin{equation}
Z_N \approx A~\mu^N~N^{\gamma-1}.
\end{equation}
Here, the factor $\mu^N$ dominates; it depends on the lattice, e.g. $\mu \approx 4.68404$
for the 3D cubic lattice. The factor $N^{\gamma-1}$ is a relatively small correction to this,
but knowledge of the exponent $\gamma$ is very useful since it is a lattice-independent (universal) exponent.
A straightforward attack on the problem
that generates all SAWs can only reach a limited
length (currently about $N=24$ for the 3D cubic lattice),
because of the large number of SAWs.
For most lattices, the value of the \emph{connective constant} $\mu$ is known only in approximation,
with a few exceptions such as the 
honeycomb lattice in 2D, with $\mu = \sqrt{2+ \sqrt{2}}$~\cite{duminil12}.
For all regular two-dimensional lattices, the exponent $\gamma$ is believed (but not proven) 
to be $\gamma_{2D}=43/32$~\cite{nienhuis};
its value in three dimensions is not known exactly
and is estimated at $\gamma_{3D} \approx 1.157$.

The history of counting SAWs goes back at least six decades,
to a paper by Orr~\cite{orr47} from 1947,
who gave the counts $Z_N, N= 1, \ldots, 6,$ for the 3D cubic lattice.
The number of steps in an enumeration
for this lattice was successively increased by 
Fisher and Sykes~\cite{fisher59},
Guttmann~\cite{guttmann87,guttmann89}, 
MacDonald et al.~\cite{macdonald92,macdonald00},
and Clisby, Liang, and Slade~\cite{clisby07}, who reached $N=30$.
Recently, we further increased the number of steps to $N=36$
by the length-doubling method~\cite{schram11}, see Section~\ref{sec:lengthdoubling},
giving $Z_{36} = 2,941,370,856,334,701,726,560,670$.
For the 2D square lattice, the current record is 
held by Jensen~\cite{jensen04},
with $Z_{71}=4,190,893,020,903,935,054,619,120,005,916$.
For more detail on many aspects of the SAW problem, see the monograph
by Madras and Slade~\cite{madras96}. 

The main goal of this article is to present an algorithm
and its implementation for counting SAWs
on a regular lattice, which is based on the length-doubling method~\cite{schram11}
we have published previously. Essentially, this method counts the number of SAWs of the double
length $2N$ by taking statistics from
the $2^N$ subsets of sites visited by each SAW of length $N$, thereby reducing the
computational effort from $\mathcal{O}\left( \mu^{2N}\right)$ to $\mathcal{O}\left((2\mu)^N\right)$.
We also discuss the use of symmetry to speed up the computation,
and the use of parallelism.
Our presentation is accompanied by a computer program 
SAWdoubler, available from \url{http://www.staff.science.uu.nl/~bisse101/SAW/}
under the GNU LGPL license.
The program can in principle handle any regular lattice, and provides
a sample implementation for the 3D cubic lattice.
It is relatively straightforward to adapt the program to other lattices,
by replacing the program file with functions
specifying the lattice,
while keeping the file with all the counting functions and data structures
unchanged. For brevity and ease of illustration,
we will often use examples from the 2D square lattice in this article.

\subsection{Length-doubling method}
\label{sec:lengthdoubling}

\begin{figure}
\begin{center}
\includegraphics[scale=0.75]{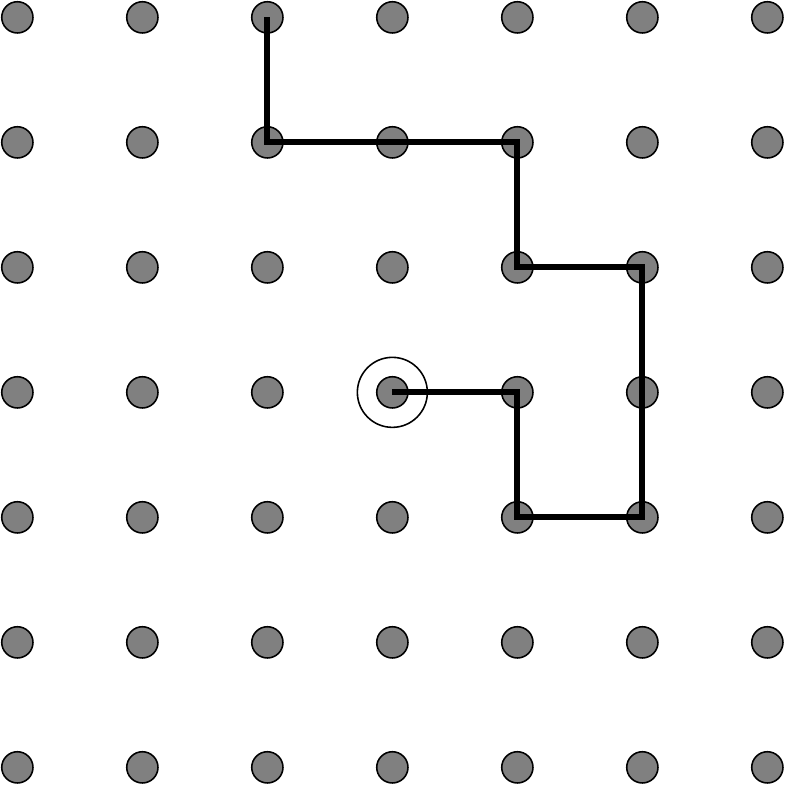}
\end{center}
\caption{Self-avoiding walk of length $N=10$ on the 2D square lattice.
The walk starts in the origin (middle of the picture).}
\label{fig:walk2d_10}
\end{figure}

A SAW of length $N$ on a regular lattice starting in the origin
can be written as a sequence
$w = (\vecr_0, \vecr_1, \ldots, \vecr_N)$ with $\vecr_0 = \vec{0}$,
meaning that we walk from the origin $\vecr_0$ to lattice
site $\vecr_1$, and so on, until we reach the end point $\vecr_N$.
Figure~\ref{fig:walk2d_10} illustrates a walk of length 10
on the square lattice in 2D.

The length-doubling method is based on combining two walks
of length $N$ into one walk of length $2N$.
Let $w, w'$ be two SAWs. We can start a walk from the end point
$\vecr_N$ of $w$
in the reverse direction of $w$ towards the origin and then continue to walk
in the direction of the end point $\vecr_N^{\prime}$ of $w'$.
This yields a walk of length $2N$. If we translate the resulting walk by 
$-\vecr_N$, we have a walk of length $2N$ starting in the origin.

The result of combining two SAWs in this way may be self-avoiding or not, depending 
on the presence of an intersection point $\vecr$.
Let $A_{\vecr}$ be the set of pairs of SAWs $(w,w')$
that both pass through the lattice point $\vecr$.
Then 
\begin{equation}
\label{eqn:z2na}
Z_{2N} = Z_N^2 - \left| \bigcup_{\vecr} A_{\vecr} \right|,
\end{equation}
because every pair $(w,w')$ of SAWs of length $N$
can be used to construct a SAW of length $2N$, except if they both pass
through a lattice point $\vecr$.
Applying the inclusion--exclusion principle from combinatorics~\cite{vanlint92}
to compute the number of elements of a union of sets from their intersections
yields the length-doubling formula
\begin{equation}
\label{eqn:doubling}
Z_{2N} = Z_N^2 + \sum_{S \neq \emptyset} (-1)^{|S|} Z_N^2(S),
\end{equation}
where $S$ is a subset of the lattice points and $Z_N(S)$
the number of SAWs that pass through all elements of $S$. 
The numbers $Z_N(S)$ can be obtained by creating all SAWs of length $N$
(but not those of length $2N$) and maintaining a bookkeeping
of all the possible sets $S$ encountered and their number of SAWs $Z_N(S)$.

The implementation of the length-doubling method poses two main challenges.
First, all sets $S$ have to be generated and, because of their large
number, be stored efficiently or only part of the sets should be stored
at the same time;
the data structure used in our implementation
is discussed in Sec.~\ref{sec:storage}. Second, the summation over these sets as given
in Eq.~(\ref{eqn:doubling}) has to be performed;
this is discussed in Sec.~\ref{sec:algo}.
We then also pay attention to how symmetry properties of the SAWs can be
exploited in Sec.~\ref{sec:symmetry}.
Our implementation SAWdoubler is tested with respect to time and memory scaling
in Sec.~\ref{sec:experiments}.
We draw conclusions and discuss future extensions in Sec.~\ref{sec:conclusions}.

\section{Storing self-avoiding walks}
\label{sec:storage}
Since all SAWs start at the origin, 
we do not need to store the starting point.
Furthermore, since the length-doubling method only cares about
whether walks of length $N$ intersect, the order of the 
sites visited in a walk is irrelevant. 
A walk can therefore be written as a set 
\begin{equation}
W = \{\vecr_1, \ldots, \vecr_N \}.
\end{equation}
Note that the same set of points $W$ can result from several different SAWs.

\subsection{Numbering the lattice sites}
The number of lattice sites that can be reached by a SAW of length $N$
is finite, and hence the sites can be numbered by a finite
numbering $\phi$,
irrespective of the dimensionality of the lattice.
For the 2D square lattice, for instance, only the $2N^2+2N+1$ points
$\vecr = (x,y)$
with $0 \leq |x|+|y| \leq N$ can be reached. 
A suitable numbering could be $0, \ldots, 2N^2+2N$.

The \emph{canonical numbering} $\phi_{\mathrm{canon}}$ for the 2D square lattice
is defined by site number $s = \phi_{\mathrm{canon}}(x,y) =(x+N)L+y+N$, 
where $L=2N+1$ is the width of the smallest square lattice enclosing all
reachable points.
This leads to a numbering 
$0, \ldots, L^2-1$, where not all sites are reachable.
In section~\ref{sec:symmetry}, a different numbering will be 
presented which facilitates exploitation of symmetry.
Using a numbering, a walk to be stored can be concisely represented by
\begin{equation}
W = \{w_1, \ldots, w_N \}, \quad \mathrm{with}~ w_1 < w_2 < \cdots < w_N.
\end{equation}
Note that the sites of $W$ are now ordered by increasing site number, and not 
by the order in which the sites are visited.

\subsection{Tree data structure}
Our aim is to store all SAWs of length $N$ in a data structure
that requires as little memory as possible, but still enables operations
such as finding all subsets $S$ of a particular walk $W$. 
We could store all SAWs simply as lists of length $N$, but this would cause
a lot of repetition, since SAWs are often similar to each other.

\begin{figure}
\begin{center}
\includegraphics[scale=0.75]{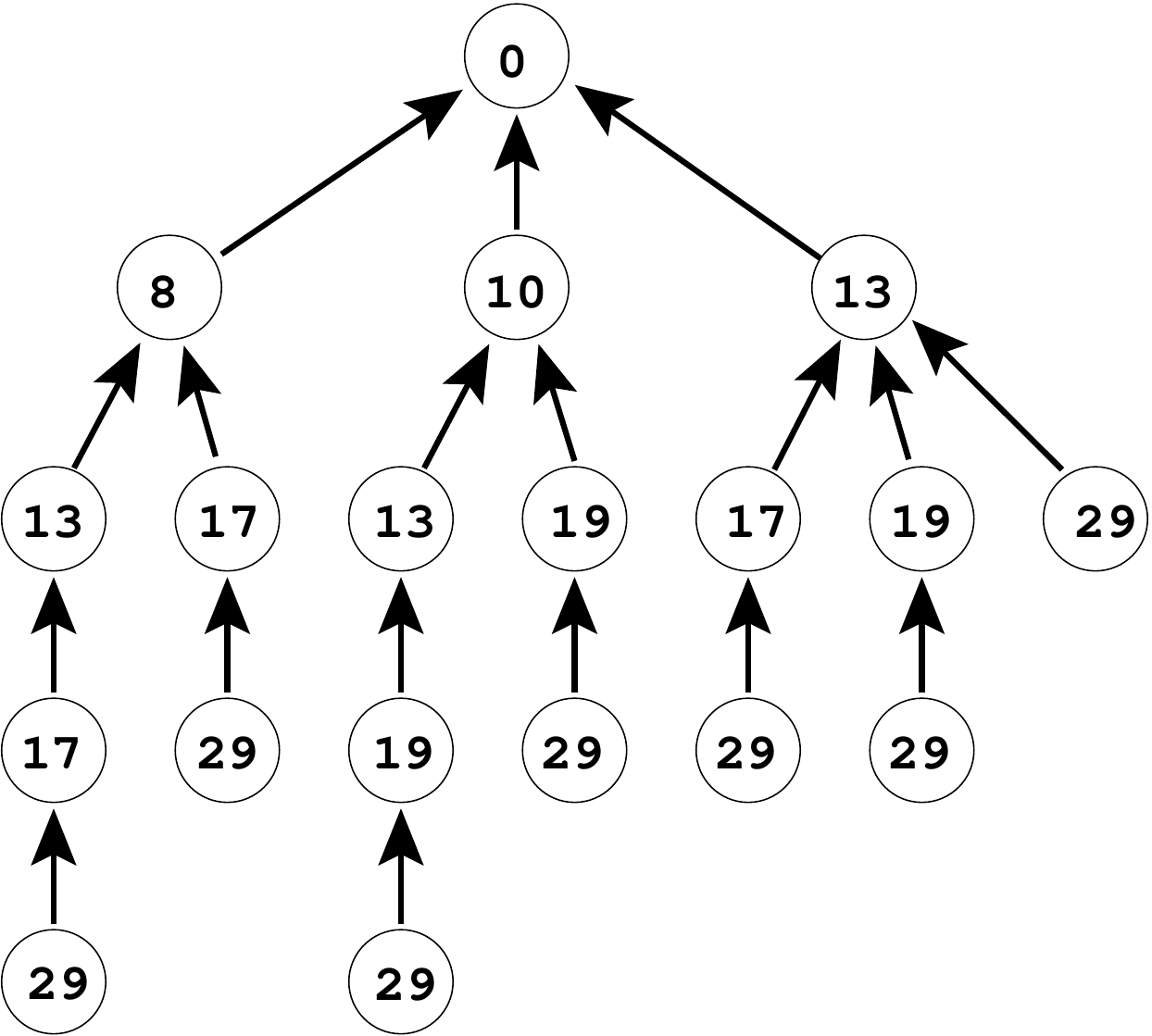}
\end{center}
\caption{Tree data structure as used by the SAWdoubler program
for storing self-avoiding walks of length $N=4$
on the 2D square lattice. 
Each tree node points to its parent. The root node is denoted by 0. 
The site numbering is the same as in Fig.~\ref{fig:symmnumbering}.
The walk $w$ from $(0,0)$ through $(0,1)$, $(-1,1)$, $(-1,0)$, to $(-2,0)$
corresponds to the walk $w=(0, 8, 17, 13, 29)$ in this site numbering,
and is stored as the set $W= \{8, 13, 17, 29 \}$ in the tree.
This tree is used for the computation of $Z_N(S)$ for all sets of sites 
$S$ that have 29 as their highest site number.
Only walks that contain 29 are stored, and only their sites $s \leq 29$.
}
\label{fig:tree}
\end{figure}

We choose a tree as our data structure, with a special extra site as the root,
with sites as nodes, and with parent--child relations defined by 
\begin{equation}
w_i = \mathit{parent}(w_{i+1}), \quad 1 \leq i < N, 
\end{equation}
for each walk $W = \{w_1, \ldots, w_N \}$.
The parent of $w_1$ is the root.
This tree data structure is illustrated by Fig.~\ref{fig:tree}.
Note that the same site number may occur several times in the tree.
The tree is constructed by consecutively adding the SAWs to be stored, 
each time checking whether the lower numbered part $w_1, \ldots, w_i$
already exists in the tree when adding site $w_i$.
If so, no new nodes need to be added 
for this part. Only when the new walk deviates from the tree, 
new nodes are introduced for the remainder $w_i, \ldots, w_N$ of the walk.

At every node of the tree, the following information is stored:
\begin{itemize}
\item $\mathit{site}$, site number of the node;
\item $\mathit{count}$, number of SAWs with this node as highest site;
\item $\mathit{child}$, first child;
\item $\mathit{sibling}$, next sibling;
\item $\mathit{parent}$, parent;
\item $\mathit{stamp}$, a time stamp (not used while building the tree);
\item $\mathit{next}$, next node with the same site number (not used while building the tree).
\end{itemize}
The variable $\mathit{site}$ can be stored using a
standard (32-bit) integer,
as site numbers remain small, growing for instance as $\mathcal{O}(N^2)$
for the 2D square lattice.
The variable $\mathit{count}$ initially (i.e.,
immediately after building the tree)
contains the count of the number of SAWs with this node as highest site.
If all walks have the same length $N$, the initial $\mathit{count}$
is nonzero only at the leaves of the tree.
The initial counts are modified during the computation
by adding counts together
so that the largest counts in the tree thus
may become of order $\mathcal{O}(Z_N)$.
Therefore, the variable $\mathit{count}$ needs a 64-bit integer.
Different walks visiting the same sites,
but in a different order, will have the same set $W$,
and hence the initial count can be larger than one.
To enable storing walks of different lengths in the same tree,
the variable $\mathit{count}$ is also present in nonleaf nodes.
In the length-doubling method, counts $Z_N(S)$ are squared,
cf. Eqn.~(\ref{eqn:doubling}), and hence a few extra long
(128-bit) integer variables such as $Z_{2N}$ must be used in order
to match the size of the counts, but such variables are not needed
in the nodes of the tree.

The variables $\mathit{child}$, $\mathit{sibling}$, and $\mathit{parent}$
are needed for traversing the tree. They point to other nodes in the tree
and are set to a dummy if no respective child, sibling, or parent exists.
Finding the parent of a node is an immediate $\mathcal{O}(1)$ operation.
Finding all children requires finding the first child using 
$\mathit{child}$, and then following the linked list of siblings
implemented by $\mathit{sibling}$. In our implementation,
the siblings are ordered by increasing site number,
which yields a rather modest savings in computation time
when processing a new sibling.
The savings are obtained in case the new sibling is already 
present in the sibling list;
otherwise, the list has to be searched until the end.
Ordering by increasing site number gives preference to lower-numbered
sites,
and these are closer to the origin and hence have more likely
been encountered already.

Two variables $\mathit{stamp}$ and $\mathit{next}$ are added to the node 
to facilitate operations of the 
counting algorithm, Algorithm~\ref{algo:correct},
see Section~\ref{sec:correction}.
The variable $\mathit{stamp}$ represents a time stamp, 
which records when we pass a certain node while traversing the tree
in the counting algorithm. 
This variable needs a 64-bit integer for storage.
Sometimes, we need to connect a set of nodes in the tree
with the same site number $s$ into a linked list. This list is implemented using
the variable $\mathit{next}$.
The total required storage per node is one 32-bit integer
and six 64-bit integers,
which amounts to 52 bytes per tree node.

The variables $\mathit{count}$, $\mathit{stamp}$, and $\mathit{next}$
may change during the counting algorithm, 
but the tree structure as defined by 
$\mathit{site}$, $\mathit{child}$, $\mathit{sibling}$, $\mathit{parent}$
remains the same after the tree has been built by 
the SAW-creating algorithm, Algorithm~\ref{algo:go},
see Section~\ref{sec:go}.
After the tree has been created, we will only use the variable $\mathit{parent}$,
and the variables $\mathit{child}$ and $\mathit{sibling}$
are not used any more; in contrast, $\mathit{stamp}$ and $\mathit{next}$
are not used during creation of the SAWs.
Therefore, some space can be saved by 
storing $\mathit{child}$ and $\mathit{stamp}$ in one field,
and the same for $\mathit{sibling}$  and $\mathit{next}$.
In our exposition, we will use the original field names,
but in our program SAWdoubler, we save the memory
of two 64-bit integers per node, reducing the required size for the tree
to 36 bytes per node. 

The width and the depth of the tree are influenced by the numbering
of the lattice sites. A careful numbering will limit
the number of children of each node, especially near the root, 
and this will enhance the reuse of initial parts of walks in the tree.
A suitable way to do this is to number the sites by increasing Euclidean
distance from the origin. For the 2D square lattice,
this limits the number of children of the root to four,
whereas an arbitrary numbering could have a much 
larger number of children and hence would lead to little reuse.

\section{Algorithms}
\label{sec:algo}
\subsection{Creating self-avoiding walks of length $N$}
\label{sec:go}
Algorithm~\ref{algo:go} gives the function Go, which creates all SAWs of length $N$
by recursively exploring all unvisited adjacent lattice sites of the current
site $\vecr_i$. When a SAW of length $N$ has been created, it is 
converted to site numbers, sorted in increasing order, and inserted into the tree
data structure.
The walk $(\vecr_0, \vecr_1, \ldots, \vecr_i)$ is stored in the array $R$,
with $R[j] = \vecr_j$ for $j=0, \ldots , i$.
The initial call of the function is 
Go$(0, N, R, \mathit{visited}, \mathit{Tree})$,
where the whole array $\mathit{visited}$
has been initialised to false, and the tree 
contains only the special root node.

\begin{algorithm}
\caption{Recursive algorithm for creating all $Z_N$ walks of length $N$}\label{algo:go}
\begin{algorithmic}[1]
\Function{Go}{$i, N, R, \mathit{visited}, \mathit{Tree}$}
   \Comment{Extend $(\vecr_0, \vecr_1, \ldots, \vecr_i)$ to length $N$}
   \State $\mathit{visited}(\vecr_i ) \gets \mathit{TRUE}$
   \If{$i=N$}
      \For{$j = 1~\mathbf{to}~ N$} 
          \State $w_j \gets \phi (\vecr_j)$ \Comment{apply numbering}
      \EndFor
      \State Sort$(W, N)$\Comment{sort in increasing order}
      \State Insert$(W,\mathit{Tree})$ 
   \Else
       \ForAll{$\vecr \in \mathit{Adj}(\vecr_i)$} \Comment{visit all neighbours of $\vecr_i$}
          \If{$\mathbf{not}~\mathit{visited}(\vecr )$}
              \State $\vecr_{i+1} \gets \vecr$ 
              \State Go$(i+1, N, R, \mathit{visited}, \mathit{Tree})$
          \EndIf
      \EndFor
   \EndIf
   \State $\mathit{visited}(\vecr_i ) \gets \mathit{FALSE}$
\EndFunction
\end{algorithmic}
\end{algorithm}

\subsection{Calculating correction terms}
\label{sec:correction}
Algorithm~\ref{algo:correct} gives the function Correct,
which calculates all correction terms 
$(-1)^{|S|}Z_N(S)^2$ of SAWs of length $N$ passing through a set $S$ of lattices sites,
by recursively expanding the set $S$ to a superset $S'$.
The initial call of the function is
Correct($\mathit{Tree}, N, \emptyset, \mathit{Bins},0$),
where $\mathit{Tree}$ has been filled by Algorithm~\ref{algo:go}
with all SAWs of length $N$.
The algorithm works as follows.

To expand the current set $S$, the algorithm first finds 
the maximum site number $s_{\max}$ for the active tree nodes.
A tree node is called \emph{active} if its walk count
contributes to the computation of the current $Z_N(S)$.
To access all active nodes with the same site number,
the algorithm uses a bin data structure. This structure stores 
the active nodes with site number $s$
together in a bin $\mathit{Bins}[s]$; 
each bin is implemented as a linked list.
At the start of the whole computation, all nodes with a nonzero count
are active. Active nodes have the current $\mathit{time}$
of the algorithm as a time stamp.
Use of such a global clock makes it easy to render many nodes inactive by just
updating the time variable.
The variable $\mathit{time}$ equals the number of different sets $S$
created so far.

As a first contribution, the set $S$, which does not contain
$s_{\max}$, is expanded by smaller sites
than $s_{\max}$.
Let $v$ be an active node with site number $s_{\max}$.
If its parent $pv$ is already active, the count of $v$ must be added into that of $pv$,
in order to give the total number of walks that pass through all sites of $S$ 
and have 
the path from the root to $pv$ as their lowest-numbered part.
If the parent is not active, its count should simply be replaced by that of $v$
and it will become active.
After that, the function Correct is recursively called to handle
all supersets $S' \supsetneq  S $ with $s_{\max} \not\in S'$.
The result is added to $Z$, with a positive sign 
since the size of $S$ is unaltered, cf.
the sign $(-1)^{|S|}$ in
Eqn.~(\ref{eqn:doubling}).
Following the call, all node counts are restored to the situation 
at the start of the function, using an undoing mechanism,
details of which we omit for the sake of brevity.

As a second contribution, the set $S$ is expanded by smaller sites
than $s_{\max}$ and also $s_{\max}$ itself is included.
All walks that do not contain $s_{\max}$ must now be discarded,
which is done by incrementing the time, 
emptying the bins of active nodes,
making the parents $pv$ active, inserting 
them into bins, and stamping them with the new time.
Also here, the function Correct is recursively called,
but now the result is subtracted as the sign $(-1)^{|S|}$
has changed, due to the expansion of $S$  by one site.
In our implementation, we also use a time stamping mechanism for 
the bins, making emptying all bins a cheap operation.

Finally, we collect and sum the squares of the counts 
for the case where $s_{\max}$ is the final site added to $S$,
i.e., the site with minimum site number of $S$,
and the set $S$ is not expanded further.

\begin{algorithm}
\caption{Recursive algorithm for calculating the correction terms $(-1)^{|S|}Z_N(S)^2$
for all sets $S$}\label{algo:correct}
\begin{small}
\begin{algorithmic}[1]
\Function{Correct}{$\mathit{Tree}, N, S, \mathit{Bins}, \mathit{time}$}
   \Comment{Correction for all $S' \supseteq S$}
   \State $Z \gets 0$
   \State $s_{\max} \gets  \max \{ s : \mathit{Bins}[s] \neq \emptyset \}$
   \Comment{maximum site number}
   \If{$s_{\max} = \mathit{root}$} 
       \State \Return $Z$
   \EndIf

   \State
   \State $\triangleright$ Contribution for $S' \supsetneq  S $
          with $s_{\max} \not\in S'$
   \ForAll{$v \in \mathit{Bins}[s_{\max}]$} \Comment{all nodes in bin}
       \State $pv = \mathit{parent}(v)$
       \If{$\mathit{stamp}(pv) = \mathit{time}$} \Comment{parent is active}
           \State $\mathit{count}(pv) \gets \mathit{count}(pv) + \mathit{count}(v)$
       \Else
           \State $\mathit{count}(pv) \gets \mathit{count}(v)$
           \State Insert$(pv, \mathit{Bins})$  \Comment{insert at header in bin}
           \State $\mathit{stamp}(pv)  \gets  \mathit{time}$ 
       \EndIf
   \EndFor
   \State $Z \gets Z  + \mathit{Correct}(\mathit{Tree}, N, S,
                     \mathit{Bins},  \mathit{time})$
   \State Restore the counts 

   \State
   \State $\triangleright$ Contribution for $S' \supsetneq  S \cup \{s_{\max}\}$
   \State $\mathit{time} \gets \mathit{time} + 1$ \Comment{include $s_{\max}$ in new $S$}
   \For{$s=0~\mathbf{to}~s_{\max}-1$}
       \State $\mathit{Bins}[s] \gets \emptyset$
              \Comment{empty the bins}
   \EndFor
   \ForAll{$v \in \mathit{Bins}[s_{\max}]$}
       \State $pv = \mathit{parent}(v)$
       \State $\mathit{count}(pv) \gets \mathit{count}(v)$
       \State Insert$(pv, \mathit{Bins})$
       \State $\mathit{stamp}(pv)  \gets  \mathit{time}$ 
   \EndFor
   \State $Z \gets Z  - \mathit{Correct}(\mathit{Tree}, N, S\cup\{s_{\max}\},
             \mathit{Bins},  \mathit{time}$) 
   \State Restore the counts
   
   \State
   \State $\triangleright$ Contribution for $S' = S \cup \{s_{\max}\}$
   \ForAll{$v \in \mathit{Bins}[s_{\max}]$} \Comment{$s_{\max}$ is final site of $S$}
       \State $Z \gets Z  + \mathit{count}(v)^2$
   \EndFor

   \State \Return $Z$
\EndFunction
\end{algorithmic}
\end{small}
\end{algorithm}

\section{Exploiting symmetry}
\label{sec:symmetry}
For the 2D square lattice, the number of SAWs
that end in a point $(x,y)$ is the same as the number ending in $(-x,y)$
because of symmetry, and similarly it is the same as the number for $(x,-y)$, $(-x,-y)$,
$(y, x)$, $(-y, x)$, $(y, -x)$, and $(-y, -x)$.
Thus, we have 8-fold symmetry which we should exploit for an efficient computation
of $Z_N$. For the 3D cubic lattice, the potential gain is even larger, 
since we have 48-fold symmetry,
obtained by composing the 8 reflections $(\pm x, \pm y, \pm z)$ with 
the 6 permutations of the variables $x,y,z$.

The symmetry operations of a lattice form a group $G$,
where every symmetry operation $Q \in G$ has an inverse symmetry operation $Q^{-1} \in G$, 
and where there is an identity operation $I \in G$, and the operations are associative.
In general, the group need not be commutative.
We denote the \emph{order}, i.e. the number of elements, of group
$G$ by $g = |G|$.
For the 2D square lattice, the group is isomorphic to the group
of \emph{signed $2 \times 2$ permutation matrices},
and its order is 8.

For a given lattice point $\vecr$, the symmetry operations that leave it invariant
form a subgroup $H_{\vecr}$ of $G$, defined by
\begin{equation}
H_{\vecr} = \{ Q \in G ~:~ Q \vecr = \vecr \}.
\end{equation}
By Lagrange's theorem~\cite{armstrong88}, the order $h_{\vecr}$
of the subgroup divides the order $g$ of $G$.
Furthermore, the symmetry number of $\vecr$, defined as 
\begin{equation}
\label{eqn:symm_r}
\mathit{Symm}({\vecr}) = |\{ Q \vecr ~:~ Q \in G \}|,
\end{equation}
satisfies 
\begin{equation}
\label{eqn:symmnr}
\mathit{Symm}({\vecr}) \cdot h_{\vecr} = g.
\end{equation}
Thus, the symmetry number of a lattice point for the 2D square lattice 
is a divisor of $g=8$. For the 3D cubic lattice, it is a divisor of 48;
this means that up to 48 different lattice points can be obtained 
by symmetry operations executed on $\vecr$. We call these points 
\emph{symmetrically equivalent} or, for short, \emph{equivalent}.
Together, these points form an equivalence class 
\begin{equation}
[ \vecr ] = \{ Q \vecr ~:~ Q \in G \}.
\end{equation}

\begin{figure}
\begin{center}
\includegraphics[scale=0.75]{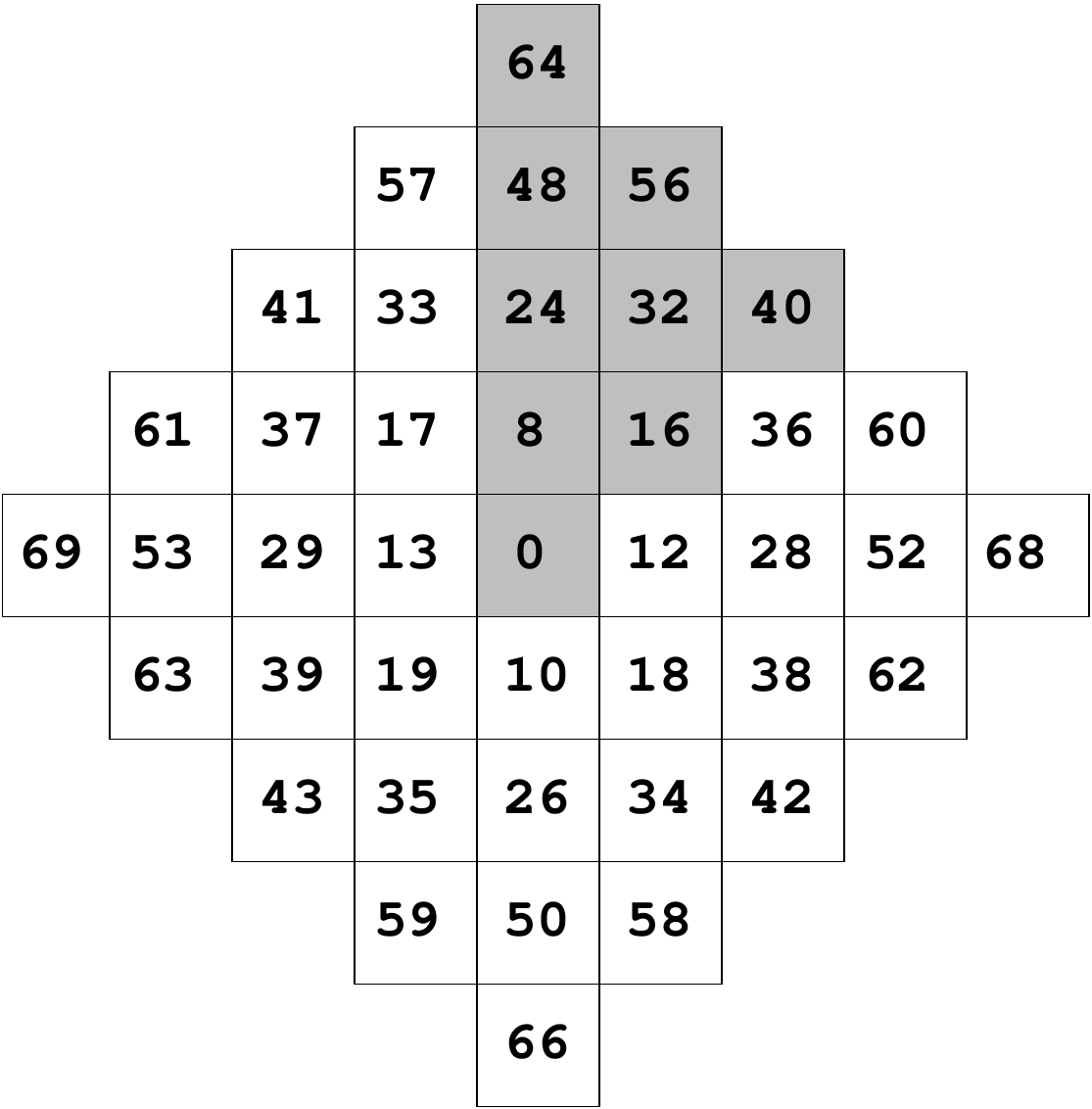}
\end{center}
\caption{Site numbering of the 2D square lattice
for self-avoiding walks of length $N=4$ as produced by the SAWdoubler
numbering function. Only reachable sites are numbered.
Lattice points $(x,y)$ with $0 \leq x \leq y$ (the grey area) are numbered first in their equivalence
class, and their site numbers are a multiple of $g=8$. 
Numbering of these points is by increasing the Euclidean distance from $(0,0)$.
Lattice point $(1,2)$ has site number 32 and symmetry number $\mathit{Symm}(32) =8$.
Its equivalence class consists of the sites 32--39.
Lattice point $(0,1)$ has site number 8 and symmetry number $\mathit{Symm}(8) =4$.
Its equivalence class consists of the sites 8, 10, 12, 13.
There are no sites 9, 11, 14, 15 in this numbering.
}
\label{fig:symmnumbering}
\end{figure}

To exploit the symmetry, the numbering 
should make it easy to determine whether two lattice points are equivalent.
This can be achieved by numbering the points from the same equivalence
class within a range of $g$ numbers, from $kg$ to $(k+1)g -1$,
for a certain $k$. There may be less than $g$ numbers 
from the range that are actually used. 
To check whether sites $s$ and $s'$ are equivalent,
we just need to divide by $g$ and round down:
\begin{equation}
s \sim s' \quad \Longleftrightarrow \quad
\left \lfloor \frac{s}{g} \right \rfloor = 
\left \lfloor \frac{s'}{g} \right \rfloor .
\end{equation}
Figure~\ref{fig:symmnumbering} shows a numbering 
that respects the symmetry for the 2D square lattice.

Let $Qs$ denote the site obtained from site $s$ by applying symmetry operation $Q$,
and $QS$ the set of sites obtained from set $S$ by applying $Q$ to the sites of $S$.
Note that $|QS| = |S|$, because $Q$ is a bijection.
Similar to Eqn.~(\ref{eqn:symm_r}) for a single lattice point,
we can define the symmetry number of a set of sites $S$,
\begin{equation}
\mathit{Symm}(S) = |\{ QS ~:~ Q \in G \}|.
\end{equation}
We can order sets of the same size lexicographically,
by comparing the highest site numbers first.
For example, the set $\{2, 4, 7\}$ is lexicographically smaller than
$\{3, 5, 7\}$, because we first compare the highest sites and find that $7 = 7$,
and then we find that $4 < 5$. 
We denote this by $\{2, 4, 7\} <_{\mathrm{lex}} \{3, 5, 7\}$.

Our aim is to compute $Z_N(S)$ for every subset $S$ of lattice sites
that occurs in a walk of length $N$.
Let $S = \{ s_1, \ldots, s_n\}$ be such a subset, with $s_1 < s_2 < \ldots < s_n$ 
and $1 \leq n \leq N$. We call the highest site $s_n$ the 
\emph{terminal site} of $S$.
Note that this is not necessarily the end point of a walk through $S$.
We can write 
\begin{equation}
S = (S \backslash \hat{S}) \cup \hat{S},
\end{equation}
where
\begin{equation}
\hat{S} = \{s \in S ~:~ s \sim s_n\}
\end{equation}
is the \emph{terminal part} of $S$,
which is the set of sites equivalent to its terminal site.
Checking whether $s \in S$ belongs to $\hat{S}$,
reduces to checking whether 
\begin{equation}
s \geq \left \lfloor \frac{s_n}{g} \right \rfloor g.
\end{equation}
It is easy to prove that
\begin{equation}
QS = Q(S \backslash \hat{S}) \cup Q\hat{S},
\end{equation}
as a disjoint union, and that
\begin{equation}
\widehat{QS} = Q\hat{S},
\end{equation}
for all subsets $S$ and all $Q \in G$.

For a set of sites $S$, we can find an operation $Q_S \in G$
such that $Q_S\hat{S}$ is lexicographically the largest
among the sets $Q\hat{S}$. The operation $Q_S$ is not unique,
but the set $Q_S\hat{S}$ is.
Since $Z_N(S) = Z_N(Q_SS)$, we need not compute $Z_N(S)$,
but we can compute $Z_N(Q_SS)$ instead.
This means that we only have to compute $Z_N(S)$
for sets $S$ with $\hat{S} \geq_{\mathrm{lex}} Q\hat{S}$ for all 
$Q \in G$. 

For every lexicographically largest set $\hat{S}$,
there are $\mathit{Symm}(\hat{S})$ symmetry operations
$Q \in G$ that lead to different sets $Q\hat{S}$.
These operations also give different sets $QS$,
because their terminal parts $\widehat{QS} = Q\hat{S}$
are different. 
Note that for the same reason,
we have 
\begin{equation}
\mathit{Symm}(\hat{S}) \leq \mathit{Symm}(S).
\end{equation}
We now just have to multiply
$Z_N(S)$ by the symmetry number
$\mathit{Symm}({\hat{S}})$ to account for all the omitted sets $S$; 
this symmetry number is most easily computed by using
\begin{equation}
\mathit{Symm}(\hat{S}) = \frac{g}{h_{\hat{S}}},
\end{equation}
similar to Eqn.~(\ref{eqn:symmnr}).

This method fully exploits the available symmetry of $\hat{S}$,
and asymptotically for large $N$ this will approach 
the full symmetry of $S$,
as in most cases the terminal part will only contain 
one site and it will have the maximum symmetry number, $g$.
Furthermore, the overhead caused by computing the symmetry number
is limited, as we only need to compute it for every $\hat{S}$,
and not for every $S$. When expanding $S$,
i.e. when adding a new, smaller site $s_{\max}$, we compute the symmetry number
if we leave the equivalence class of the terminal site 
(i.e., $s_{\max} \not\sim s_n$).
From then onwards, $\hat{S}$ cannot change anymore,
and we use its value for all $S$ with the same terminal part $\hat{S}$.
It would also be possible to exploit the full symmetry of $S$ instead
of only that of $\hat{S}$,
but this would yield only limited gain
and would cause some extra overhead.  

\section{Experimental results}
\label{sec:experiments}

In this section, we will test the performance of the SAWdoubler program
both with respect to computation time and memory.
In our previous work~\cite{schram11}, we used 200 processing cores of a supercomputer 
and spent about 50,000 core hours for the computation of $Z_{36}$ for the 3D cubic lattice.
In the present work, we will focus instead on the 
performance on a PC with a limited
amount of memory. Our test case is the same 3D cubic lattice.

The test architecture we use is a dual-core Apple MacBook Pro with a 2.53 GHz Intel Core i5
dual-core processor and 4 GB RAM, a 256 KB L2-cache per core, a 3 MB L3-cache,
and a 5400 rpm hard disk of size 500 GB, running the MacOs 10.6.8
operating system.
We use the gcc compiler, version 4.2, with flags \texttt{-O3 -Wall}. 

SAWdoubler first creates SAWs of length $N$ by Algorithm~\ref{algo:go}
and then computes the correction terms 
$(-1)^{|S|} Z_N(S)$ by Algorithm~\ref{algo:correct}.
The computation of values $Z_N(S)$ for sets $S$ has been organised
such that all sets $S$ with the same terminal site $t$ are handled by a separate tree.
A SAW $w$ of length $N$ is then only stored if it contains $t$, 
and only the sites $s \leq t$ of the walk
are stored in the tree; thus stored walks may have length less than $N$.
The main program then loops over $t$ up to the maximum reachable lattice site.
This procedure saves much memory, and makes it possible to reach larger $N$.
We call this the \emph{split-tree} approach.
If desired, the single site $t$ could be replaced by a set of sites $T$,
to reduce memory requirements further.

Table~\ref{tbl:time} presents the computation time needed for calculating
$Z_{2N}$, for $N=7, \ldots, 14$. 
The time given is the total elapsed time of a single run
as measured by the Unix \texttt{time} utility. (For $N \leq 6$,
the time needed is too short and our measurement 
becomes inaccurate; therefore we omit those results.)
In almost all cases, the elapsed time is close to the used CPU time.
Comparing columns in the table without and with symmetry shows 
that exploiting symmetry considerably accelerates the computation,
by up to a factor of 26.2 for $N=14$.

We use two different numberings in our experiments for Table~\ref{tbl:time}.
Changing the numbering from ordering by the Euclidean norm to ordering by the 
Manhattan norm ($||\vecx||_1 = \sum_i |x_i|$), 
saves up to a modest 5 per cent in time for $N \leq 13$
but it takes about 10 per cent more memory. 
This becomes a disadvantage for $N=14$, where the amount of memory required is
close to the total amount available.
Both numberings order the lattice points by an increasing distance from the origin, 
given by the respective norm, and thus perform much better than
other numberings (that we used in our initial implementations of SAWdoubler.)

The last column of Table~\ref{tbl:time} represents an attempt to use
the full computing capability of the dual-core architecture by employing both 
cores in parallel. This is done by running two instances
of the program simultaneously, one executing the odd iterations of the main loop,
and the other the even ones. This already gives a speedup of around 1.7
for $N=9$--$13$ and the $L_2$ norm. The load imbalance here is reasonable, 
with the largest imbalance (3.5 per cent above the average time) observed 
for $N=13$, one core running 687 s and the other 736 s. 
Both cores use the same shared memory,
so they may hinder each other and both must store a complete tree in memory.
For $N=14$, the trees become very large, and together they fill up about two thirds of
the available RAM memory. Here, the CPU time was about 10 per cent
less than the elapsed time, 
perhaps caused by cores interfering with each other
when making use of shared resources such as the RAM and the L3 cache.
This difference between CPU time and elapsed time only occurred for the
largest problem instance $N=14$. 
The resulting speedup for $N=14$ is 1.56 out of 2.

The dual-core approach can be generalised to more cores by cyclic assignment:
processor core $c=0,\ldots, p-1$ from a set of $p$ cores will carry out 
iterations $c, c+p, c+2p, \ldots $ of the main loop.
For larger $N$ and a larger number of cores $p$,
this static distribution of work by cyclic assignment may lead to 
larger imbalance than that observed for two cores,
in particular since the amount of work may then vary considerably
between loop iterations. In that case, a dynamic distribution 
of work based on a job queue would lead to better balance.

Considering the growth of the computation time with increasing $N$,
we note that moving from $N=12$ to $N=13$ increases the time by a factor of 7.1
(using symmetry and the $L_2$ norm), 
and moving from $N=13$ to $N=14$ by a factor of 7.4.
Asymptotically, the length-doubling method grows as 
$\mathcal{O}((2\mu)^N) \approx \mathcal{O}(9.368^N)$,
since every one of the $Z_N = \mathcal{O}(\mu^N)$ walks of length $N$ has
$2^N$ different subsets $S$ and incrementing a counter for each of these
costs $\mathcal{O}((2\mu)^N)$ operations.
The memory savings of the tree by eliminating repetition 
also pays off in computation time, as counters may now be incremented 
by larger numbers than one. This results in slower initial growth
than the factor of 9.368 theoretically predicted.

For comparison, we also used our program
to compute $Z_N$ in a straighforward way, without length-doubling
and without using symmetry, just by creating and counting all $Z_N$ walks.
The computation of $Z_{17}$ in this manner already took 9258 s,
about the same time as the computation of $Z_{28}$ with length-doubling
and symmetry.

Examining the breakdown of the computation time, we observed that for large
$N$ by far most of the time is spent in computing the correction terms
by Algorithm~\ref{algo:correct}. 
A notable 25 per cent of that time is spent in finding the largest 
remaining site $s_{\max}$ using the bin structure, 
and the remainder in traversing the tree. 
Finding $s_{\max}$ can possibly be optimised in the future, 
perhaps by using some form of hashing, as many bins will be empty.

\begin{table}
\begin{tabular}{rr@{.}lr@{.}lr@{.}lr@{.}l}
 \hline
$N$ & \multicolumn{2}{c}{No symmetry}
    & \multicolumn{2}{c}{Symmetry}
    & \multicolumn{2}{c}{Symmetry}
    & \multicolumn{2}{c}{Symmetry}\\
    & \multicolumn{2}{c}{$L_2$ norm}
    & \multicolumn{2}{c}{$L_2$ norm}
    & \multicolumn{2}{c}{$L_1$ norm}
    & \multicolumn{2}{c}{$L_2$ norm}\\
    & \multicolumn{2}{c}{1 core} 
    & \multicolumn{2}{c}{1 core} 
    & \multicolumn{2}{c}{1 core} 
    & \multicolumn{2}{c}{2 cores}\\
 \hline
 \hline
7 & 0&210 & 0&016 &0&020 &0&017\\
8 & 1&42 & 0&091&0&089 &0&062\\
9 & 9&70 &0&534 &0&518 &0&316\\
10 & 69&8 &3&49 &3&35 &1&96\\
11 &530& &24&5 &23&3&14&1 \\
12 &4110& &177&&169& &102&\\
13 & 30990& &1259& &1213& &736&\\
14 &244235& &9331 &&9417& &5976&\\ 
  \hline
\end{tabular}
\caption{Time (in s) of the computation of the number of SAWs of length $2N$
for the 3D cubic lattice. Exploitation of 48-fold symmetry 
is either disabled or enabled; 
the site numbering is based on either the 
$L_2$ (Euclidean) norm or the $L_1$ (Manhattan) norm;
and either one or two cores of the dual-core processor are used.}
\label{tbl:time}
\end{table}

Table~\ref{tbl:memory} presents the memory requirements of the SAWdoubler program
for the split-tree approach, and for comparison also for the approach 
where the tree is not split, i.e. the single-tree approach.
These requirements should be compared with $Z_N$, the number of SAWs $w$
of length $N$,
and also with the related number $Z_N^{\prime}$ of sets of sites $W$
obtained from the SAWs, i.e. ignoring the walk order;
the value of $Z_N^{\prime}$ does not depend on the chosen numbering.
It holds trivially that $Z_N^{\prime} \leq Z_N$.
Comparing the single-tree storage with its lower bound $Z_N^{\prime}$
and its upper bound $NZ_N$, we observe that the storage is within
a range of 1.82--2.09 times the lower bound,
and that it is far from the upper bound.
Using a tree thus saves a lot of memory.
The number $Z_N^{\prime}$ is easily obtained
by counting the leaves of the single tree,
as each set $W$ must have its own leaf in the tree data structure.

Splitting the tree, by only storing walks 
with a particular terminal site $t$, further reduces memory, 
by up to a factor 109 for $N=11$, and makes computations possible
for $N=12, 13, 14$ that are otherwise infeasible.
Memory usage can be reduced
by another factor of at least 1.4 by deleting the terminal site itself
from the tree;
this means deleting one layer of leaves, e.g. deleting
node 29 everywhere in Fig.~\ref{fig:tree}.
Since we use 36 bytes per node for storing the tree itself,
and another 16 bytes per node for the undoing mechanism,
we need a total of 2.0 GB storage for $N=14$.

\begin{table}
\begin{tabular}{rrrrrrr}
 \hline
$N$ & \multicolumn{1}{c}{$Z_N$}
    & \multicolumn{1}{c}{$N \cdot Z_N$}
    & \multicolumn{1}{c}{Nodes}
    & \multicolumn{1}{c}{Leaves}
    & \multicolumn{1}{c}{Max nodes}
    & \multicolumn{1}{c}{Max nodes} \\
    &       &               & \multicolumn{1}{c}{single tree} 
    & \multicolumn{1}{c}{single tree} 
    & \multicolumn{1}{c}{split tree} & \multicolumn{1}{c}{split tree}\\
    &&&& $= Z_N^{\prime}$
    & \multicolumn{1}{c}{with term.}& \multicolumn{1}{c}{w/o term.}\\
 \hline
 \hline
7 & 81\,390& 569\,730& 129\,846& 71\,498&2\,672 &1\,692 \\
8 & 387\,966& 3\,103\,728 &643\,824&341\,421& 11\,927 &7\,886\\
9 & 1\,853\,886& 16\,684\,974& 3\,150\,431& 1\,601\,052&38\,205 &25\,981\\
10 & 8\,809\,878 &88\,098\,780 & 15\,367\,644 &7\,596\,096&183\,532 &122\,983\\
11 & 41\,934\,150& 461\,275\,650& 74\,587\,922&35\,616\,048 &682\,590 &465\,637 \\
12 & 198\,842\,742 & 2\,386\,112\,904 &---& ---&2\,854\,104 &1\,969\,834\\
13 & 943\,974\,510& 12\,271\,668\,630&---& ---&11\,961\,303 &8\,234\,139 \\
14 & 4\,468\,911\,678& 62\,564\,763\,492&---&---& 54\,177\,636&37\,849\,701\\
  \hline
\end{tabular}
\caption{Memory usage (in number of nodes) during
the computation of the count $Z_{2N}$ of SAWs of length $2N$
for the 3D cubic lattice. The value $N \cdot Z_N$ is the storage 
needed if every SAW of length $N$ would be stored separately in an array
of length $N$. The single-tree columns give the storage (number of nodes) for
one tree storing all the SAWs of length $N$ and also the number of leaves.
The split-tree columns give the maximum number of nodes of a tree in case 
a separate tree is used for all sets $S$ with the same terminal site,
with or without the terminal site stored.
Site numbering is by the Euclidean norm.}
\label{tbl:memory}
\end{table}

\section{Conclusion and future work}
\label{sec:conclusions}

In this article, we have presented an algorithm 
for counting the number of self-avoiding walks of length $2N$
by creating self-avoiding walks of length $N$,
based on the length-doubling method~\cite{schram11}.
We have made available a program SAWdoubler in C, which implements the 
method, exploits symmetry, and uses an efficient data structure.

We have computed $Z_{28} = 12, 198,184,788,179,866,902 $ for the 3D cubic lattice
on a dual-core laptop computer with 4 GB main memory in 1 hour and 40 minutes, 
and thereby demonstrated the efficiency of our program.
We have verified the counting results up to $N=28$ of our 
previous work~\cite{schram11}, which was done by a completely different implementation.
Furthermore, we have shown that two processor cores of a dual-core processor
can be used with a speedup
of 1.7, provided two copies of the problem tree fit into the shared memory. 

The design of the SAWdoubler program makes it easy to extend the computation
to other lattices. Anyone can replace the file \texttt{lattice.c} (aimed at the 3D cubic lattice)
by another file such as for the 2D square or honeycomb lattice,
the 3D BCC or FCC lattice, or the 4D hypercubic lattice, which is straightforward to do,
and no change in the tree structure functions of the file \texttt{sawdoubler.c} is needed,
nor changes in the main file \texttt{saw.c}.

For future work, the software can be extended to compute $Z_{2N+1}$ as well as $Z_{2N}$,
and to compute squared end-to-end distances $||\vecr_N - \vecr_0||^2$, 
as has been done in~\cite{schram11}.
A limitation of the present software is the size of the tree
for one terminal site $t$. Generalising to a terminal set $T$ to keep the tree
within any amount of available memory would be the next step.
Future research could investigate variants of the present problem,
such as self-avoiding polygons and lattices with forbidden regions.
The present work should provide an efficient and extendible basis for such 
investigations.
%\end{linenumbers}


\begin{thebibliography}{99}

\bibitem{duminil12}
H. Duminil-Copin and S. Smirnov,
{\it The connective constant of the honeycomb lattice equals $\sqrt{2+\sqrt2}$},
Ann. of Math. {\bf 175}, 1653--1665 (2012).

\bibitem{nienhuis}
B. Nienhuis, 
{\it Exact critical point and exponents of the $O(n)$ model in
two dimensions}, Phys. Rev. Lett. {\bf 49}, 1062--1065 (1982).

\bibitem{orr47}
W. J. C. Orr,
{\it Statistical treatment of polymer solutions at infinite dilution},
Trans. Faraday Soc. {\bf 43}, 12--27 (1947).

\bibitem{fisher59}
M. E. Fisher and M. F. Sykes,
{\it Excluded-volume problem and the Ising model of ferromagnetism},
Phys. Rev. {\bf 114}, 45--58 (1959).

\bibitem{guttmann87}
A. J. Guttmann,
{\it On the critical behaviour of self-avoiding walks},
J. Phys. A: Math. Gen. {\bf 20}, 1839--1854 (1987).

\bibitem{guttmann89}
A. J. Guttmann,
{\it On the critical behaviour of self-avoiding walks: II},
J. Phys. A: Math. Gen. {\bf 22}, 2807--2813 (1989).

\bibitem{macdonald92}
D. MacDonald, D. L. Hunter, K. Kelly, and N. Jan,
{\it Self-avoiding walks in two to five dimensions:
exact enumerations and series study},
J. Phys. A: Math. Gen. {\bf 25}, 1429--1440 (1992).

\bibitem{macdonald00}
D. MacDonald, S. Joseph, D. L. Hunter, L. L. Moseley, N. Jan,
and A. J. Guttmann,
{\it Self-avoiding walks on the simple cubic lattice},
J. Phys. A: Math. Gen. {\bf 33}, 5973--5983 (2000).

\bibitem{clisby07}
N. Clisby, R. Liang, and G. Slade, 
{\it Self-avoiding walk enumeration via the lace expansion},
J. Phys. A: Math. Theor. {\bf 40}, 10973--11017 (2007).

\bibitem{schram11}
R. D. Schram, G. T. Barkema, and R. H. Bisseling,
{\it Exact enumeration of self-avoiding walks},
J. Stat. Mech.,  p06019 (2011).

\bibitem{jensen04}
I. Jensen,
{\it Enumeration of self-avoiding walks on the square lattice},
J. Phys. A: Math. Gen. {\bf 37}, 5503--5524 (2004).

\bibitem{madras96}
N. Madras and G. Slade, {\it The Self-Avoiding Walk}, (Birkh\"{a}user, Boston, 1993).

\bibitem{vanlint92}
J.H. van Lint and R.M. Wilson,
{\it A Course in Combinatorics},
(Cambridge University Press, Cambridge, UK, 1992).

\bibitem{armstrong88}
M. A. Armstrong,
{\it Groups and Symmetry},
(Springer, New York, 1988).


\end{thebibliography}
\end{document}